\documentstyle[11pt]{article}
\begin{document}


\def\a{\alpha}
\def\b{\beta}
\def\d{\delta}
\def\e{\epsilon}
\def\g{\gamma}
\def\k{\kappa}
\def\l{\lambda}
\def\o{\omega}
\def\t{\theta}
\def\s{\sigma}
\def\z{{\bar{z}}}
\def\D{\Delta}
\def\L{\Lambda}
\def\T{\Theta}

\def\p{\partial}
\def\cp2{\cos\frac{\phi}{2}}

\def\cL{{\cal L}}
\def\cB{{\cal B}}


\def\beq{\begin{equation}}
\def\eeq{\end{equation}}
\def\bea{\begin{eqnarray}}
\def\eea{\end{eqnarray}}
\def\ba{\begin{array}}
\def\ea{\end{array}}
\def\no{\nonumber}
\def\lt{\left}
\def\rt{\right}
\newcommand{\bq}{\begin{quote}}
\newcommand{\eq}{\end{quote}}

\begin{titlepage}
\begin{flushright}
YITP/K-1109\\
DPSU-95-2\\
hep-th/9506157
\end{flushright}
\vskip.3in
\begin{center}
{\huge Supersymmetric Extension of the Sine-Gordon Theory with Integrable
Boundary Interactions}
\vskip.3in
{\Large Takeo Inami$~^a$,
Satoru Odake$~^{b,}$\footnote{
Email: odake@yukawa.kyoto-u.ac.jp }}
and
{\Large Yao-Zhong Zhang$~^{a,}$\footnote{
Email: yzzhang@yukawa.kyoto-u.ac.jp; yzz@maths.uq.oz.au.}}
\vskip.1in
$a$. Yukawa Institute for Theoretical
Physics, Kyoto University, Kyoto 606, Japan

$b$. Department of Physics, Faculty of Science, Shinshu University,
Matsumoto 390, Japan

\end{center}
\vskip.4in
\begin{center}
{\bf Abstract}
\end{center}
\begin{quote}
Integrability and supersymmetry of the supersymmetric extension
of the sine-Gordon theory on a
half-line are examined and the boundary potential which preserves both the
integrability and supersymmetry  on the bulk is derived. It appears that
unlike the boundary
bosonic sine-Gordon theory, integrability and supersymmetry
strongly restrict the form and parameters of the boundary potential, so that
no free parameter in the boundary term is allowed up to a choice of signs.

\end{quote}

\end{titlepage}
\newpage

\newcommand{\sect}[1]{\setcounter{equation}{0}\section{#1}}
\renewcommand{\theequation}{\thesection.\arabic{equation}}

\sect{Introduction\label{intro}}

Integrable quantum field theories in 1+1 dimensions, both massless and
massive, have had interesting applications in particle physics as
well as  statistical physics near criticality. Many of physical systems
in nature have boundaries. In view of applications to such systems,
including open string field theory and dissipative quantum mechanics, attempts
have recently been made for extending conformal field
theories and massive integrable quantum field theories to incorporate
boundary interactions, in  both field theoretical
\cite{Che84,Car89,Gho94,Cor94} and algebraic
\cite{Skl88,Mez91} approaches.

A particularly interesting model of massive boundary field theories is the
sine-Gordon theory on a half-line \cite{Skl87}.
The boundary potential has been obtained
and the associated boundary $S$-matrix been constructed by Ghoshal and
Zamolodchikov in \cite{Gho94}. One may
take the $y$-direction
to be the euclidean time and $x$-direction to be the space. The
euclidean action of the boundary sine-Gordon theory has been shown
to be of the form
\beq
S=\int^\infty_{-\infty} dy\int^0_{-\infty}dx\lt \{
  \frac{1}{2}\left(\p_x\phi\right)^2+\frac{1}{2}\left(\p_y\phi\right )^2
  -\frac{m^2}{\b^2}\cos\b\phi\rt\}+
  \L\int^\infty_{-\infty}dy\;\lt .\cos\frac{\b(\phi-\phi_0)}{2}\rt |_{x=0},
\eeq
where $\phi(x,y)$ is a real scalar field, $\b$ is a dimensionless
coupling constant and $m$ is the mass parameter representing the deviation
from the massless theory.
This model is integrable for arbitrary choices of the
parameters $\L$ and $\phi_0$ \cite{Gho94}.  The same type
of analysis has been performed for the (real coupling) affine Toda theories
by Corrigan et al \cite{Cor94}.

In this paper we study supersymmetric integrable quantum
field theories with boundaries. Supersymmetry appears to play a significant
role in physics, e.g. superstring field theory,
quantum spin-1 chain  and surface
roughening-reconstruction competition.
We consider supersymmetric extension of the sine-Gordon theory
on a half-line and derive a boundary potential which preserves both
integrability and supersymmetry on the bulk.
The bulk supersymmetric sine-Gordon (SSG) theory can be regarded
as integrable deformation of the supersymmetric free field theory (with
central charge $c=3/2$). The boundary conditions for the $c=3/2$ theory
has previously been studied in the context of open superstring
theory \cite{DiV82}. Only the Neumann
boundary condition has been considered, however.

We will first examine the integrability condition and find a few different
types of integrable boundary terms. Supersymmetry is broken in general in
a theory with boundaries unless suitable boundary conditions are chosen.
We will derive
a general form of boundary potential which restores (one half of)
supersymmetry. We will then compare the result from integrability with that
from supersymmetry.

\sect{SSG Theory on the Bulk\label{whole-line}}

The euclidean action of the SSG theory
on the whole line $(-\infty,\infty)$ or on the bulk is given by
\bea
S_0&=&\int dx\,dy\;\cL_0,\no\\
{\cal L}_0&=&\frac{1}{2}\left(\p_x\phi\right)^2+\frac{1}{2}\left(\p_y\phi
  \right)^2-\bar{\psi}(\p_x-i\p_y)\bar{\psi}+\psi(\p_x+i\p_y)\psi\no\\
& &  -\frac{m^2}{\b^2}\cos\b\phi-2m\bar{\psi}
   \psi\cos\frac{\b\phi}{2},\label{S-0}
\eea
where $\psi,~\bar{\psi}$ are the two
components of a Majorana fermion.

We use the notation: $z=x+iy,~\bar{z}=x-iy$.
The classical equations of motion on the bulk are
\bea
&&\partial_{\bar{z}}\p_z\phi=\frac{m^2}{4\b}\sin\b\phi+\frac{m\b}{4}
   \bar{\psi}\psi\sin
   \frac{\b\phi}{2},\no\\
&&\p_{\bar{z}}\psi=-\frac{m}{2}\bar{\psi}\cos\frac{\b\phi}{2},\no\\
&&\p_z\bar{\psi}=-\frac{m}{2}\psi\cos\frac{\b\phi}{2}.\label{bulk-eqm}
\eea
The SSG theory on the whole line  is integrable.
It is the simplest of supersymmetric affine Toda theory, and
is known to be  associated with
the superalgebra $osp(2|2)^{(2)}$. An infinite number of
conserved charges has been derived at the classical level
\cite{Gir78} and  the quantum level \cite{Sas88}.

The supersymmetry of the model is manifest if we write the above component
expression in terms of superfield:
\beq
S_0=\int d^2z\,d^2\t\;\lt\{\frac{1}{2}\bar{D}\Phi D\Phi-
    \frac{2m}{\b^2}\cos\frac{\b\Phi}{2}\rt \},
\eeq
where $\bar{D}$ and $D$ are covariant derivatives in the superspace coordinates
$\z,~\bar{\t}$ and $z,~\t$, respectively,
\beq
\bar{D}=-\p_{\bar{\t}}+\bar{\t}\p_\z,~~~~~~
   D=\p_\t+\t\p_z
\eeq
and $\Phi$ is a (scalar) superfield,
\beq
\Phi=\phi+i\t\psi-i\bar{\t}\bar{\psi}+i\bar{\t}\t F
\eeq
with $F$ an auxiliary field. The equation of motion in terms of the
superfield reads
\beq
\bar{D}D\Phi=\frac{m}{\b}\sin\frac{\b\Phi}{2}.
\eeq

\sect{SSG Theory on a Half-Line\label{integrability}}

The theory on a half-line $x\in (-\infty, 0]$ is defined by adding the boundary
term to the bulk part $S_0$ of the action,
\beq
S=S_0+S_b\equiv \int_{-\infty}^\infty dy\int_{-\infty}^0dx\;{\cal L}_0
  +\int_{-\infty}^\infty dy\; {\cal B}(\phi,\psi,\bar{\psi})
\eeq
where the boundary potential
$\cB$ is assumed to be a functional of the fields at $x=0$ but not
their derivatives.
In addition to the bulk field equation
(\ref{bulk-eqm}), we have equations of motion at the boundary $x=0$,
\bea
\p_x\phi+\frac{\p\cB}{\p\phi}&=&0,\no\\
\psi-\frac{\p\cB}{\p\psi}&=&0,\no\\
\bar{\psi}+\frac{\p\cB}{\p\bar{\psi}}&=&0.\label{boundary-condition}
\eea
{}From this, we remark that at $x=0$,
$\p^2\cB/ \p\phi\p\psi
=\p^2\cB/ \p\phi\p\bar{\psi}
=\p^2\cB/ \p\psi\p\bar{\psi}=0$.

For simplicity we rescale the fields $\phi,~\psi$ and $\bar{\psi}$ in the
following way: $\phi\rightarrow \phi/ \b,~\psi\rightarrow \psi/ \b,~
\bar{\psi}\rightarrow\bar{\psi}/ \b$, and restrict our attention to
the classical case $\b\rightarrow 0$.
Moreover we  set $m=2$ so that the bulk equations of motion simplify to
the form
\bea
&&\partial_{\bar{z}}\p_z\phi=\sin\phi+\frac{1}{2}
   \bar{\psi}\psi\sin
   \frac{\phi}{2},\no\\
&&\p_{\bar{z}}\psi=-\bar{\psi}\cos\frac{\phi}{2},\no\\
&&\p_z\bar{\psi}=-\psi\cos\frac{\phi}{2}.\label{s-bulk-eqm}
\eea

As is known \cite{Gir78,Sas88},
in the bulk theory there is an infinite number of conserved
charges constructed from densities $T_{s+1},~\bar{T}_{s+1},~
\T_{s-1}$ and $\bar{\T}_{s-1}$ with $s=1,3,5,
\cdots$. These densities satisfy
the following continuity equations
\beq
\p_\z T_{s+1}=\p_z\T_{s-1}\,,~~~~~~\p_z\bar{T}_{s+1}=\p_\z\bar{\T}_{s-1}.
\eeq
The densities of $s=1$  are given by the energy-momentum tensor,
\bea
&&T_2=\lt (\p_z\phi\rt )^2-\p_z\psi\psi,\no\\
&&\bar{T}_2=\lt (\p_\z\phi\rt )^2+\p_\z\bar{\psi}\bar{\psi},\no\\
&&\T_0=\bar{\T}_0=-2\cos\phi-\bar{\psi}\psi\cp2.
\eea
The $s=3$ densities are \cite{Gir78,Sas88}
\bea
T_4&=&\lt (\p_z^2\phi\rt )^2-\frac{1}{4}\lt (\p_z\phi\rt )^4+\frac{3}{4}
     \lt (\p_z\phi\rt )^2\p_z\psi\psi-\p_z^2\psi\p_z\psi,\no\\
\bar{T}_4&=&\lt (\p_\z^2\phi\rt )^2-\frac{1}{4}\lt (
     \p_\z\phi\rt )^4-\frac{3}{4}
     \lt (\p_\z\phi\rt )^2\p_\z\bar{\psi}\bar{\psi}+
     \p_\z^2\bar{\psi}\p_\z\bar{\psi},\no\\
\T_2&=&\lt (\p_z\phi\rt )^2\cos\phi-\p_z\psi\psi\cos^2\frac{\phi}{2}
     -\bar{\psi}\p_z\psi\p_z\cos\frac{\phi}{2}+\frac{1}{4}\bar{\psi}\psi
     \lt (\p_z\phi\rt )^2\cos\frac{\phi}{2},\no\\
\bar{\T}_2&=&\lt (\p_\z\phi\rt )^2\cos\phi+\p_\z\bar{\psi}
     \bar{\psi}\cos^2\frac{\phi}{2}
     +\psi\p_\z\bar{\psi}\p_\z\cos\frac{\phi}{2}+
     \frac{1}{4}\bar{\psi}\psi
     \lt (\p_\z\phi\rt )^2\cos\frac{\phi}{2}.\label{theta-thetabar}
\eea

Suppose that one can choose boundary potential such that at $x=0$
\beq
T_4-\bar{T}_4-(\T_2-\bar{\T}_2)=\frac{d}{dy}\Sigma_3(y),\label{sigma}
\eeq
where $\Sigma_3(y)$ is some functional of boundary fields.
Then the charge $P_3$, given by
\beq
P_3=\int^0_{-\infty}dx\;(
T_4+\bar{T}_4+\T_2+\bar{\T}_2)-i\Sigma_3(y)
\eeq
is a non-trivial integral of motion.

We now examine in what circumstances  $T_4-\bar{T}_4-(\T_2-\bar{\T}_2)$ may
be written as a total $y$-derivative. Observing that
\bea
\psi_{xx}&=&4\psi\cos^2\frac{\phi}{2}-2i\bar{\psi}_y\cos\frac{\phi}{2}
   -2\bar{\psi}\p_x\cp2-i\psi_{xy},\no\\
\bar{\psi}_{xx}&=&4\bar{\psi}\cos^2\frac{\phi}{2}+2i\psi_y\cos\frac{\phi}{2}
   -2\psi\p_x\cp2+i\bar{\psi}_{xy},\no\\
\psi_{xy}&=&-2\bar{\psi}_y\cp2-2\bar{\psi}\p_y\cp2-i\psi_{yy},\no\\
\bar{\psi}_{xy}&=&-2\psi_y\cp2-2\psi\p_y\cp2+i\bar{\psi}_{yy},
\eea
one can show that
\bea
T_4-\bar{T}_4-(\T_2-\bar{\T}_2)&=&-\frac{i}{2}(\phi_{xx}-\phi_{yy})
   \phi_{xy}+\frac{i}{8}(\phi_x^2-\phi_y^2)\phi_x\phi_y+i\phi_x
   \phi_y\cos\phi\no\\
& &+\frac{3i}{16}(\phi_x^2-\phi_y^2)(\bar{\psi}_y\bar{\psi}-\psi_y\psi)
   +i3\cos^2\frac{\phi}{2}(\bar{\psi}_y\bar{\psi}-\psi_y\psi)\no\\
& &-i\p_x\cos\frac{\phi}{2}(\bar{\psi}_y\psi-\psi_y\bar{\psi})
   +i(\bar{\psi}_{yy}\bar{\psi}_y-\psi_{yy}\psi_y)\no\\
& &+i\bar{\psi}\psi\phi_x\phi_y\cos\frac{\phi}{2}-\frac{3}{8}
   \phi_x\phi_y(\psi_y\psi+\bar{\psi}_y\bar{\psi})\no\\
& &+2\p_y\cos\frac{\phi}{2}(\bar{\psi}_y\psi-\bar{\psi}\psi_y)
   +\cos\frac{\phi}{2}(\bar{\psi}\psi_{yy}-\bar{\psi}_{yy}\psi),
   \label{derivative1}
\eea
where  $Q_x,~Q_{xy}$ etc stand for $\p_xQ,~\p_x\p_yQ$ and so on.
If one notices that
\bea
&&\phi_{xx}=4\sin\phi-\phi_{yy}+2\bar{\psi}\psi\sin\frac{\phi}{2},\no\\
&&\phi_{xy}=-\frac{\p^2\cB}{\p\phi^2}\phi_y-\frac{\p^2\cB}{\p\psi\p\phi}
  \psi_y-\frac{\p^2\cB}{\p\bar{\psi}\p\phi}\bar{\psi}_y
  =-\frac{\p^2\cB}{\p\phi^2}\phi_y,
\eea
eq.(\ref{derivative1}) can be written as
\bea
T_4+\bar{\T}_2-\bar{T}_4-\T_2&=&(``{\rm bosonic~part}")+
    i\bar{\psi}\psi\phi_y\lt (\frac{\p^2\cB}{\p\phi^2}\sin\frac{\phi}{2}
    -\frac{\p\cB}{\p\phi}\cos\frac{\phi}{2}\rt ) \no\\
& &+\frac{3i}{16}\lt (\lt (\frac{\p\cB}{\p\phi}\rt )^2-\phi_y^2
    +16\cos^2\frac{\phi}{2}\rt )(\bar{\psi}_y\bar{\psi}-\psi_y\psi)\no\\
& &-i\p_x\cos\frac{\phi}{2}(\bar{\psi}_y\psi-\psi_y\bar{\psi})
   +i(\bar{\psi}_{yy}\bar{\psi}_y-\psi_{yy}\psi_y)\no\\
& &+\frac{3}{8}
   \frac{\p\cB}{\p\phi}\phi_y(\psi_y\psi+\bar{\psi}_y\bar{\psi})\no\\
& &+2\p_y\cos\frac{\phi}{2}(\bar{\psi}_y\psi-\bar{\psi}\psi_y)
   +\cos\frac{\phi}{2}(\bar{\psi}\psi_{yy}-\bar{\psi}_{yy}\psi),
   \label{derivative2}
\eea
where the ``bosonic part" is given by
\bea
  &&
  -i\frac{\p^2\cB}{\p\phi^2}\phi_{yy}\phi_y
  +\frac{i}{8}\frac{\p\cB}{\p\phi}\phi_y^3
  +i\biggl(-\frac{1}{8}\Bigl(\frac{\p\cB}{\p\phi}\Bigr)^3
    +2\frac{\p^2\cB}{\p\phi^2}\sin\phi-\frac{\p\cB}{\p\phi}\cos\phi
  \biggr)\phi_y \no\\
  &&~~~~~~~~=
  i\partial_y\Biggl(\frac{1}{8}\cB\phi_y^2
  +\int^{\phi}\biggl(
    -\frac{1}{8}\Bigl(\frac{\p\cB}{\p\phi}\Bigr)^3
    +2\frac{\p^2\cB}{\p\phi^2}\sin\phi-\frac{\p\cB}{\p\phi}\cos\phi
  \biggr)d\phi\Biggr) \no\\
  &&~~~~~~~~~~~
  -i\Bigl(\frac{\p^2\cB}{\p\phi^2}+\frac{1}{4}\cB\Bigr)\phi_y\phi_{yy}
  +\frac{i}{8}\phi_y^2(\bar{\psi}_y\bar{\psi}-\psi_y\psi).
  \label{bosonic-part}
\eea

Taking account of the result in the bosonic case \cite{Gho94}, we look for the
solution $\cB(\phi,\psi,\bar{\psi})$ of the form
\beq
\cB(\phi,\psi,\bar{\psi})=\cB_b(\phi)+\cB_f(\psi,\bar{\psi}),
\eeq
where $\cB_b$ is the boundary potential in the bosonic sine-Gordon theory:
$\cB_b=\L\cos\frac{\phi-\phi_0}{2}$ with $\L$ and $\phi_0$ being arbitrary
constants; $\cB_f$ is a
function of the fields $\psi$ and $\bar{\psi}$. The purely bosonic
part in (\ref{derivative2}) is then a total $y$-derivative, and hence we only
have to deal with the remaining bilinear terms in $\psi$ and $\bar{\psi}$.
$\cB_f$ can be written as
\beq
\cB_f(\psi,\bar{\psi})=M\bar{\psi}\psi
   +\e\psi+\bar{\e}\bar{\psi}
\eeq
with $M$ being a bosonic parameter and
$\e,~\bar{\e}$
constant fermionic parameters.
Then the equations of motion at boundary $x=0$ become
\bea
  1)&&
  \psi=-\frac{\e+M\bar\e}{1-M^2},\quad
  \bar\psi=\frac{\bar\e+M\e}{1-M^2}\qquad \mbox{for }M\neq\pm1,\\
  2)&&
  \bar\psi=\mp(\psi+\e),\quad \bar\e=\mp\e\qquad\qquad\quad
  \mbox{for }M=\pm1.
\eea
It is not difficult to see, after some exercise, that we have two solutions
in order for the r.h.s. of
(\ref{derivative2}) to be a total $y$-derivative
\bea
&& 1)~~~M\neq \pm 1,~\L,~\phi_0,~\e,~\bar{\e}~{\rm arbitrary};\\
&& 2)~~~M=\pm 1,~~~~~\L=\pm 8,~~~~~\phi_0=0,~~~~~\bar{\e}=0=\e.
\eea
Returning to the original normalization, one has two different forms of
boundary potential compatible with
the integrability on the bulk:
\bea
&& 1)~~~\cB(\phi,\psi,\bar{\psi})=\L\cos\frac{\b(\phi-\phi_0)}{2}
        +M\bar{\psi}\psi+\e\psi+\bar{\e}\bar{\psi},~~(M\neq\pm 1)
        ; \label{boundary-term-b}\\
&& 2)~~~\cB(\phi,\psi,\bar{\psi})=\pm \frac{4m}{\b^2}\cos\frac{\b\phi}{2}
        \pm \bar{\psi}\psi
        . \label{boundary-term1}
\eea
These equations give rise to the following boundary conditions at
$x=0$, respectively,
\bea
1)~~\p_x\phi&=&\frac{\b\L}{2}\sin\frac{\b(\phi-\phi_0)}{2},~~
     \psi=-\frac{\e+M\bar{\e}}{1-M^2},~~~
       \bar{\psi}=\frac{\bar{\e}+M\e}{1-M^2}, ~~(
       M\neq \pm 1);\label{boundary-condition-b}\\
2)~~\p_x\phi&=&\pm \frac{2m}{\b}\sin\frac{\b\phi}{2},
    ~~~~\psi\pm \bar{\psi}=0.\label{boundary-condition1}
\eea

These boundary potentials have been derived
by examining the first non-trivial conserved
charge $P_3$, where we have restricted attention to the classical case.
We believe that our analysis will be completed by showing
that all conserved charges of higher spin give the same results, and
that the above
computation can be extended to the quantum theory by taking into account
the normal ordering.

\sect{Supersymmetric Boundary Interactions
        on the Half-Line\label{supersymmetry}}

Again we work with (\ref{s-bulk-eqm}) and ${\cal L}_0~({\cal S}_0)$
corresponding to it.
Supersymmetry transformation is given by
\bea
&&\d_s\phi=\eta\psi+\bar{\eta}\bar{\psi},\no\\
&&\d_s\psi=-\eta\p_z\phi-2\bar{\eta}\sin\frac{\phi}{2},\no\\
&&\d_s\bar{\psi}=\bar{\eta}\p_{\bar{z}}\phi+2\eta\sin
   \frac{\phi}{2},\label{super-trans}
\eea
where $\eta,~\bar{\eta}$ are fermionic parameters.
It can be checked that under the transformation  (\ref{super-trans}),
${\cal L}_0$ changes by a total derivative:
\bea
\d_s{\cal L}_0&=&\p_x\left\{\frac{1}{2}\p_x\phi (\eta\psi+\bar{\eta}\bar{\psi})
   -2\sin\frac{\phi}{2}(\eta\bar{\psi}+\bar{\eta}\psi)+\frac{i}{2}\p_y\phi
   (\eta\psi-\bar{\eta}\bar{\psi})\right\}\no\\
& &+\p_y\left\{\frac{1}{2}\p_y\phi (\eta\psi+\bar{\eta}\bar{\psi})
   +i2\sin\frac{\phi}{2}(\eta\bar{\psi}-\bar{\eta}\psi)-\frac{i}{2}\p_x\phi
   (\eta\psi-\bar{\eta}\bar{\psi})\right\}.\label{total-derivative1}
\eea
It follows immediately that the theory defined by the lagrangian $\cL_0$
is supersymmetric on the whole line.
However, this is not true for the theory
on the half-line, because the boundary destroys
the supersymmetry of the action. In fact, (\ref{total-derivative1})
implies that
\beq
\d_s S_0=\int^\infty_{-\infty}dy\;\lt .\left\{\frac{1}{2}\p_x\phi
    (\eta\psi+\bar{\eta}\bar{\psi})
   -2\sin\frac{\phi}{2}(\eta\bar{\psi}+\bar{\eta}\psi)+\frac{i}{2}\p_y\phi
   (\eta\psi-\bar{\eta}\bar{\psi})\right\}\rt |_{x=0}.\label{delta-S0}
\eeq
The r.h.s. of the above equation is non-zero.

In order to preserve supersymmetry for the theory on the half-line one may
either add a boundary term to the action
or impose boundary condition by hand on $\phi,~\psi$ and $\bar{\psi}$,
in such a way as to cancel the total
$x$-derivative terms in (\ref{total-derivative1}).

The boundary potential which restores supersymmetry is obtained by solving
the  equation
\beq
\d_sS_0+\d_sS_\cB=0.
\eeq
On dimensional consideration, we take the candidate for $\cB$ to be of the form
\beq
\cB=\L_s \cos\frac{\phi-\phi_s}{2}+M_s\bar{\psi}\psi
\eeq
where $\L_s,\phi_s$ and $M_s$ are constant bosonic parameters. Then
\bea
\d_sS_\cB&=&\int^\infty_{-\infty}dy\;\lt\{-\frac{\L_s}{2}\sin\frac{\phi-\phi_s}
   {2}\d_s\phi+M_s\d_s\bar{\psi}\psi+M_s\bar{\psi}\d_s\psi\rt \}\no\\
&=&\int^\infty_{-\infty}dy\lt\{\lt (-\frac{\L_s}{2}\sin\frac{\phi-\phi_s}{2}
   +2M_s\sin\frac{\phi}{2}\rt )(\eta\psi+\bar{\eta}\bar{\psi})\rt .\no\\
& &~~~~~~~~~~~~~~~~\lt . +\frac{M_s}{2}\p_x\phi(\bar{\eta}\psi+\eta\bar{\psi})
   +\frac{iM_s}{2}\p_y\phi(\bar{\eta}\psi-\eta\bar{\psi})\rt
\}.\label{delta-Sb}
\eea
It turns out that there is no solution unless
\beq
\bar{\eta}=\mp\eta.\label{half}
\eeq
The fact that only one of $\eta$ and $\bar{\eta}$ is independent implies that
only half of the supersymmetry on the bulk is preserved on the half-line.
One encounters the same situation in the case of open superstring theory
\cite{DiV82}.

Restricting to the choice (\ref{half}), we have found two solutions:
\beq
M_s=\pm 1,~~~~~\phi_s=0,~~~~~\L_s=\pm 8.
\eeq
Therefore the boundary term which cancels the total derivative of the
lagrangian $\cL_0$ is given by
\beq
\cB=\pm \lt (8\cos\frac{\phi}{2}+\bar{\psi}\psi\rt ).\label{boundary-term2-s}
\eeq
The existence of two solutions corresponding to the choice of two kinds of
half supersymmetry (\ref{half}) is reminiscent of two models of superstring,
Neveu-Schwarz and Ramond.

It remains to examine whether the boundary potential
(\ref{boundary-term2-s}) has a room for other
terms. Such terms, if they exist, must be invariant under the (half-)
supersymmetry transformation. Allowing for fermionic parameters,
we propose the following term to be added to the
boundary potential (\ref{boundary-term2-s}):
$\e_s\psi+\bar{\e}_s\bar{\psi}$.
Under the supersymmetry transformation, it transforms as
\beq
\e_s\d_s\psi+\bar{\e}_s\d_s\bar{\psi}=-\frac{1}{2}\p_x\phi(\e_s\pm
  \bar{\e}_s)\eta+2\sin\frac{\phi}{2}(\bar{\e}_s\pm\e_s)\eta
  +i\frac{1}{2}\p_y\phi(\e_s\mp\bar{\e}_s)\eta.\label{half-trans}
\eeq
We find that (\ref{half-trans}) is a $y$-derivative if
\beq
\bar{\e}_s=\mp\e_s.
\eeq
This implies that the term $\e_s(\psi\mp\bar{\psi})$ is left invariant
under the half supersymmetry transformation and therefore can be freely
added to the boundary potential (\ref{boundary-term2-s}).

Returning to the original normalization, we have
the boundary potential which preserves the supersymmetry on the bulk:
\beq
\cB=  \pm \frac{4m}{\b^2}\cos\frac{\b\phi}{2}\pm\bar{\psi}\psi
  +\e_s(\psi\mp\bar{\psi}),\label{boundary-term2}
\eeq
which leads to the following boundary condition at $x=0$:
\bea
&&\p_x\phi=\pm \frac{2m}{\b}\sin\frac{\b\phi}{2},\no\\
&&\psi\pm \bar{\psi}=-\e_s.\label{boundary-condition2}
\eea
This boundary condition is supersymmetry preserving, i.e. for
$\bar{\eta}=\mp\eta$ we have
$\d_s\lt (\p_x\phi\mp 4\sin\frac{\phi}{2}\rt )=-i\eta\p_y(\psi\pm\bar{\psi}
   +\e_s)=0$ and
$\d_s(\psi\pm\bar{\psi}+\e_s)=-\eta\lt (\p_x\phi\mp 4\sin\frac{\phi}{2}\rt
)=0$.

Let us examine the possibility of
imposing boundary condition by hand.
It is easily seen that for arbitrary parameters $\eta$ and $\bar{\eta}$
there is no nontrivial choice of boundary condition
for the r.h.s. of eq.(\ref{delta-S0}) to be vanishing. However if one sets
$\bar{\eta}=\mp\eta$ in (\ref{delta-S0}), then
\beq
\d_s S_0=\int^\infty_{-\infty}dy\;\lt .\left\{\lt (\frac{1}{2}\p_x\phi
   \pm 2\sin\frac{\phi}{2}\rt )\eta (\psi\mp \bar{\psi})+\frac{i}{2}\p_y\phi
   \eta(\psi\pm \bar{\psi})\right\}\rt |_{x=0}\label{delta/2-S0}
\eeq
and one may have two choices of boundary conditions at $x=0$:
\bea
i)~~~&&\p_x\phi=\mp 4\sin\frac{\phi}{2},~~~~~
    \psi\pm \bar{\psi}=0;\label{boundary-condition3}\\
ii)~~~&&\p_y\phi=0,~~~~~
    \psi\mp\bar{\psi}=0.\label{boundary-condition4}
\eea
We want to see if these boundary conditions imposed by hand are left invariant
by the supersymmetry transformation $\d_s$ with the parameters
$\eta,~\bar{\eta}$ satisfying $\bar{\eta}=\mp \eta$. It is easy to check
that (\ref{boundary-condition4}) is supersymmetric invariant
whereas (\ref{boundary-condition3}) is not.

Both the boundary conditions (\ref{boundary-condition3}) and
(\ref{boundary-condition4}) do not coincide with
that in section 3. Therefore they are not compatible
with the integrability on the half-line.

\sect{Integrability vs. Supersymmetry: Conclusion\label{inte-super}}

As can be seen from sections \ref{integrability} and
\ref{supersymmetry}, only when one chooses $\e_s=0=\bar{\e}_s$,  are
(\ref{boundary-term2}) and (\ref{boundary-condition2}) derived from
supersymmetry consideration
compatible with (\ref{boundary-term1}) and (\ref{boundary-condition1})
derived from integrability consideration. Therefore the total euclidean
action which preserves both integrability and supersymmetry on the
bulk is
\beq
S=\int^\infty_{-\infty}dy\int^0_{\infty}dx{\cal
L}_0+\int^\infty_{-\infty}dy\lt.
   \lt\{\pm\frac{4m}{\b^2}\cos\frac{\b\phi}{2}\pm\bar{\psi}\psi\rt\}\rt
|_{x=0},
   \label{final-result}
\eeq
and the boundary condition at $x=0$ is
\beq
\p_x\phi=\pm \frac{2m}{\b}\sin\frac{\b\phi}{2},
    ~~~~\psi\pm \bar{\psi}=0.\label{final-boundary-condition}
\eeq

To summarize, we have constructed the boundary potential for
the supersymmetric extension of the sine-Gordon theory on the half-line by
imposing both integrability and supersymmetry. The potential thus obtained
is unique modulo overall sign. In addition there is a class of boundary
potential (\ref{boundary-term-b})
which  preserves integrability but breaks supersymmetry and
another class (\ref{boundary-term2})
which preserves supersymmetry but is not integrable.

%

\vskip.3in
We would like to thank K. Kobayashi,  R. Sasaki, K. Schoutens
and T. Uematsu for discussions.
This work was partially supported by the Grant-in-Aid
for Scientific Research from the Ministry of Education, Science and
Culture of Japan. Y.Z.Z. is supported by the Kyoto University
Foundation.

%


\end{document}